# Detecting Topological Currents in Graphene Superlattices

R. V. Gorbachev[1,2]†, J. C. W. Song[3,4]†, G. L. Yu[1], A. V. Kretinin[2], F. Withers[2], Y. Cao[1], A. Mishchenko[1], I. V. Grigorieva[2], K. S. Novoselov[2], L. S. Levitov[3*], A. K. Geim[1,2*]

[1]Centre for Mesoscience and Nanotechnology, University of Manchester, Manchester M13 9PL, UK
[2]School of Physics and Astronomy, University of Manchester, Oxford Road, Manchester, M13 9PL, UK
[3]Department of Physics, Massachusetts Institute of Technology, Cambridge, MA 02139, USA
[4]School of Engineering and Applied Sciences, Harvard University, Cambridge, Massachusetts 02138, USA

†these authors contributed equally to this work
*levitov@mit.edu; geim@man.ac.uk

**Topological materials may exhibit Hall-like currents flowing transversely to the applied electric field even in the absence of a magnetic field. In graphene superlattices, which have broken inversion symmetry, topological currents originating from graphene's two valleys are predicted to flow in opposite directions and combine to produce long-range charge neutral flow. We observe this effect as a nonlocal voltage at zero magnetic field in a narrow energy range near Dirac points at distances as large as several microns away from the nominal current path. Locally, topological currents are comparable in strength to the applied current, indicating large valley-Hall angles. The long-range character of topological currents and their transistor-like control by gate voltage can be exploited for information processing based on the valley degrees of freedom.**

Berry curvature is a physical field intrinsic to some crystal lattices, which can dramatically impact the transport properties of materials [1-6]. Topological effects, while known for some time [7], have gained attention recently in connection with the discovery of topological insulators [8-11]. In these materials, topological bands lead to new phenomena such as topologically protected edge-state transport in zero magnetic field [12-14]. No less striking, however, is the expected impact of Berry curvature on bulk transport, leading to topological currents flowing perpendicular to the applied electric field $E$ [5,6]. The nondissipative nature of these currents, ensured by their transverse character, resembles that of Hall currents. However, topological currents can arise in the absence of magnetic field $B$ and even without breaking time reversal symmetry (TRS). In contrast to cyclotron orbits in a magnetic field drifting perpendicular to $E$, topological currents originate from perfectly straight trajectories that skew left or right relative to $E$ (Fig. 1A). In materials whose electronic structure has more than one valley [4,15,16], topological currents in different valleys have opposite signs and, if intervalley scattering is weak, can add up to produce long-range topological charge-neutral currents.

Graphene placed on top of hexagonal boron nitride (G/hBN) (Fig. 1) affords a unique venue for inducing and manipulating topological bulk currents at $B=0$ for the following reasons. First, graphene's band structure possesses a nonzero Berry's phase [17], a prerequisite for the existence of Berry curvature. Second, if crystallographic axes of graphene and hBN are aligned introducing a global A/B sublattice asymmetry [18,19], inversion symmetry is no longer respected, which is key to observing Berry curvature (see below and Supplementary Information [20]). Third, the high electronic quality of graphene protects topological currents against intervalley scattering, allowing them to propagate away from the applied current path. The valley Hall effect (VHE) creates an electrical response in remote regions, which can be exploited to detect the presence of topological currents experimentally. The nonlocal response is expected whenever the Fermi level in graphene is tuned into one of Berry curvature hot spots (Fig. 1B). This approach, besides offering a direct probe of topological currents, provides a precision tool for mapping out the Berry curvature in G/hBN Bloch bands. We note that, so far, it has proven challenging to probe bulk topological currents without applying *B*. This was the case, in particular, for the anomalous Hall effect in magnetic metals, perhaps the cleanest system available previously to study topological bulk bands [6].

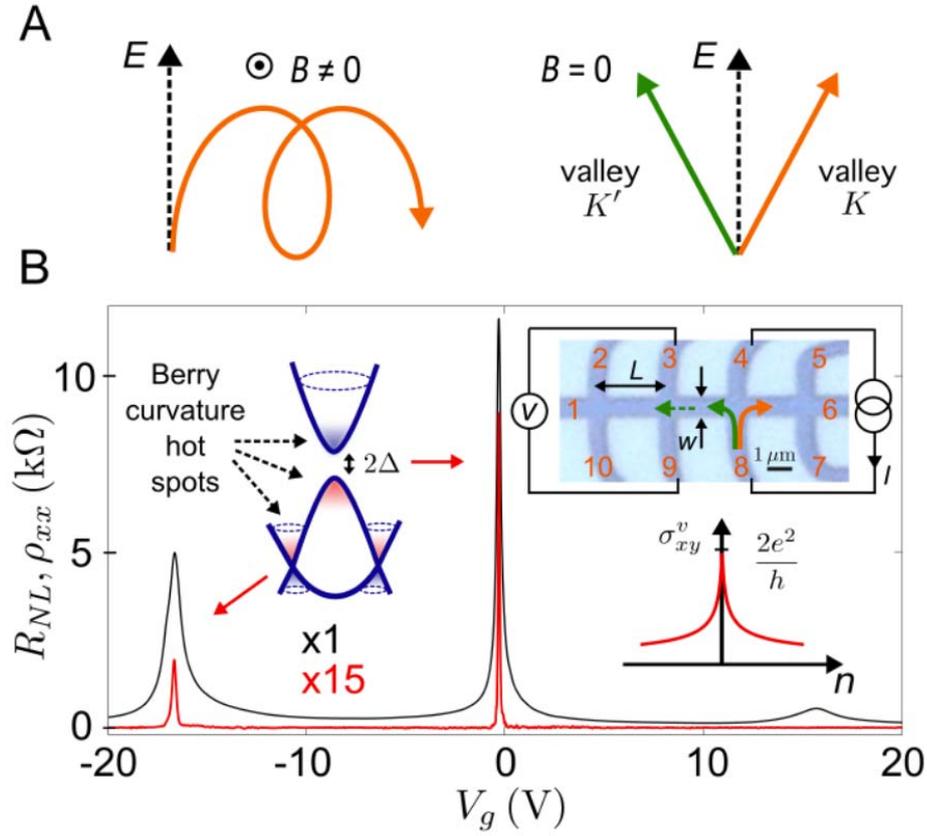

*Fig. 1. Detection of long-range valley transport due to topological currents. (A) Non-topological and topological Hall currents. Left: Drifting cyclotron orbits give rise to Hall currents of the same sign for valleys K and K'. Right: Skewed motion induced by Berry curvature. Trajectories are straight lines directed at nonzero angles to the longitudinal field, having opposite signs for valleys K and K'. The net transverse current, which is charge-neutral, creates a nonlocal charge response via a reverse VHE. (B) Nonlocal resistance in graphene superlattices (red curve) and longitudinal resistance (black curve) measured in G/hBN superlattices (graphene aligned on hBN; refs. 18-23). The back gate voltage $V_g$ was applied through a $\approx$130 nm thick dielectric (SiO$_2$ plus hBN); T =20 K. Top right inset: Optical micrograph of our typical G/hBN device and the nonlocal measurement geometry; L $\approx$3.5 μm, w $\approx$1 μm. Shown schematically are valley K and K' currents and the long-range response mechanism. Left inset: Schematic band structure of graphene superlattices with Berry curvature hot spots arising near the gap opening and avoided band crossing regions [20,25]. Bottom right inset: Valley Hall conductivity modeled for gapped Dirac fermions as a function of carrier density.*

Experimentally, we studied G/hBN superlattices (in which the graphene and hBN lattices were aligned) fabricated following the procedures described in ref. 21. Fifteen superlattice devices similar to the one shown in the inset of Fig. 1B were investigated. The charge carrier mobilities $\mu$ varied from 40,000 to 80,000 cm$^2$ V$^{-1}$ s$^{-1}$, and the longitudinal and Hall resistivities ($\rho_{xx}$ and $\rho_{xy}$, respectively) exhibited essentially the same behavior as reported previously [18,19,21,22]. Namely, pronounced peaks in $\rho_{xx}$ were observed at the main neutrality point (NP) and at carrier density $n \approx \pm 3\times10^{12}$ cm$^{-2}$ (Fig. 1B, Supplementary Fig. S1). The peaks in $\rho_{xx}$ were accompanied by a sign reversal of $\rho_{xy}$ indicating the emergence of new NPs in the valence and conduction bands of graphene [18-22]. Both encapsulated and non-encapsulated structures were investigated, with the former having an additional hBN crystal placed on top of a G/hBN superlattice to protect it from environment [20,21]. Some of the latter showed activation behavior at the main NP (Supplementary Fig. S2), yielding a bandgap $2\Delta$ of 350±40 K [20], in agreement with the earlier transport [18,19] and spectroscopy [23] measurements. Our encapsulated devices, despite higher electronic quality, exhibited no activation behavior at the main NP, with the $\rho_{xx}$ value saturating at <10 kOhm for temperatures $T$ below 50 K [19,21]. Although this behavior remains to be understood, it is likely that devices that do not show activation behavior in transport properties still have a bandgap [20]. We note in this regard that observing the activation behavior usually relies on midgap impurity states which pin the Fermi level inside the gap [20]. However, few midgap states are expected in high-quality G/hBN devices. Alternatively, spatial charge inhomogeneity can "short-circuit" the activation behavior by allowing current to circumnavigate the insulating regions. This would also obscure the activation behavior, leading to metallic-like transport in $\rho_{xx}$. However, charge inhomogeneity is expected to have relatively little effect on the VHE: being a non-sign-changing function of density (inset of Fig. 1B), it should not average out. Therefore, the activation behavior is not essential for observing topological currents [20]. Indeed, both encapsulated and non-encapsulated graphene were found to exhibit very similar nonlocal response as reported below.

The central result of our study is shown in Fig. 1B. The nonlocal resistance $R_{NL}$ was determined by applying current between, e.g., contacts 4 and 8 and measuring voltage between, e.g., 3 and 9 (micrograph of Fig. 1B). $R_{NL}$ exhibits large sharp peaks at the main and hole-side NPs (unless stated otherwise, all the presented data refer to zero $B$). A striking feature of the observed nonlocality is its narrow range in $n$ (Figs. 1B, 2A). Unlike $\rho_{xx}$, which follows the $1/n$ dependence typical for graphene and remains sizeable (>100 Ohm) over the entire range of accessible $n$ (Figs. 1,2 and Supplementary Fig. S1), $R_{NL}$ decays rapidly with $n$ and completely disappears under noise for densities >$10^{11}$ cm$^{-2}$ away from the NPs. The dependence can be approximately described by $R_{NL} \propto 1/|n|^\alpha$ with $\alpha \approx$ 2.5-3 (Fig. 2B). A nonlocal voltage can also appear due to stray charge currents, described by the van der Pauw relation [24] $R_{NL} \sim \rho_{xx} \exp(-\pi L/w)$ where $L$ is the distance between the current path and voltage probes and $w$ the device width (micrograph of Fig. 1B). For typical aspect ratios $L/w \approx 4$, the formula yields ≈0.01 Ohm. The magnitude of this contribution and its $n$ dependence, which follows $\rho_{xx}(n)$, are clearly incompatible with the observed nonlocal response.

We have also investigated how $R_{NL}$ depends on $L$ and found an exponential dependence exp(-$L/\xi$) with $\xi \approx$1.0 μm (Fig. 2C). This $\xi$ value is close to $w$ and much larger than the elastic mean free path of ≈0.1 μm estimated from $\mu$ for the range of $n$ where $R_{NL}$ appears. Furthermore, $R_{NL}$ exponentially decreases with increasing $T$ so that no nonlocal signal is detected above 150 K whereas $\rho_{xx}$ remains large at this $T$ at the main NP (Supplementary Figs. S2,S3). Non-encapsulated devices exhibited practically the same behavior of $R_{NL}$ as a function of $n$ and $T$ (Supplementary Fig. S4) but the absolute value of $R_{NL}$ was somewhat smaller than that in encapsulated devices [20]. In addition, we investigated the effect of charge inhomogeneity $\delta n$ on $R_{NL}$, which was controlled by sweeping to progressively larger gate voltages $V_g$ beyond the hole-side NP [20]. When $\delta n$ was increased by a factor of ≈2, $R_{NL}$ could sometimes change by more than an order of magnitude [20]. The inhomogeneity enhanced $R_{NL}$ at the main NP and suppressed it at the hole-side NP (Supplementary Fig. S5). The difference is attributed to a narrow energy width of the secondary Dirac spectrum [20].

To confirm the key role played by Berry curvature, we verified that the nonlocal response was absent in G/hBN systems without alignment and, accordingly, with no detectable superlattice effects [18-22]. This is illustrated in Fig. 2A that shows $R_{NL}$ for aligned and nonaligned devices with the same $\mu$ and in the same geometry. In the nonaligned devices (>20 measured), no nonlocal signal could be observed even at our maximum resolution of

~0.1 Ohm (blue curve in Fig. 2A). Therefore, the observed $R_{NL}$ cannot be explained by charge-neutral flow of spin and/or energy, which are indifferent to crystal alignment. The latter flows also require broken TRS and completely disappear in zero $B$ as reported previously [24-26]. In our superlattice devices, a contribution of spin/energy flows becomes appreciable only for $B$ >0.1 T (Supplementary Fig. S6), leading to rapid broadening of the $R_{NL}$ peaks, in agreement with ref. 24.

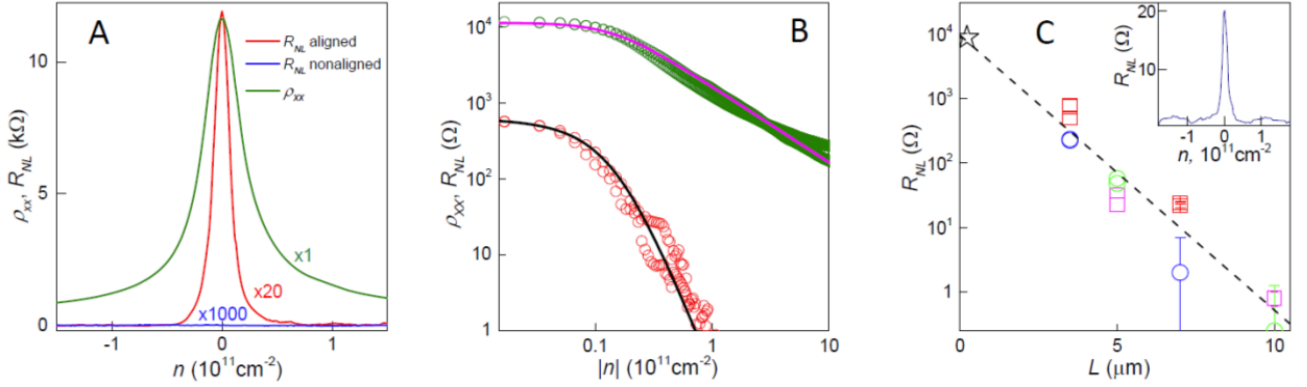

*Fig. 2. Density and distance dependences for nonlocal valley currents.* **(A)** Behavior of $\rho_{xx}$ and $R_{NL}$ near the main NP in superlattice devices (green and red, respectively). The blue curve is for a nonaligned, reference device ($\approx$10° misalignment between graphene and hBN crystal axes). $L \approx$3.5 µm; $w \approx$1 µm; $T$ =20 K. **(B)** The same data on a logarithmic scale. Away from the NP, $\rho_{xx}$ exhibits the conventional $1/n$ dependence (green) and can be described by $\rho_{xx}(n) \propto 1/(n^2+\delta n^2)^{1/2}$ where $\delta n \approx 1.5 \times 10^{10}$ cm$^{-2}$ accounts for charge inhomogeneity (magenta). Measured $R_{NL}$ (red symbols) and the Berry curvature model (black curve, Eq. 3) using the above $\delta n$ and $\Delta$ =180 K found from local measurements [18-20]. **(C)** Nonlocal signal decays exponentially with increasing $L$. Color refers to similar devices with the same $L$ or the same device with two different $L$; $w \approx$1 µm, 20 K; 10 devices in total. The black star is the anomalous contribution to $\rho_{xx}$ observed in the bend geometry, which is consistent with strong topological currents induced locally [20]. Inset: $R_{NL}$ remains sizeable at $L \approx$7 µm.

We have also considered the possibility that the observed nonlocality at $B$=0 may originate from an edge transport mechanism. For example, topological materials can support gapless edge modes that coexist with the gapped bulk. Such modes could in principle mediate nonlocal charge transport. However, our experiments provide no evidence for metallic conductivity along the device edges. First, the measured nonlocal response was similar for all our devices, independent of whether they exhibited metallic or insulating behavior in $\rho_{xx}$. Second, AFM studies show that the moiré pattern associated with the alignment extends all the way to the device edges and there are no distinct edge regions [19]. Third, extrapolating the dependence in Fig. 2C to small $L$, we find that the edge transport scenario would require metallic conductivity along edges of $\approx 2e^2/h$ over distances of $L$ >1 µm. This would imply perfect edge state transport, which is unlikely at such length scales in the presence of strong intervalley scattering expected at microscopically rough edges.

Proceeding with the analysis, we note that the aligned superlattices are comprised of hexagonal unit cells, each representing a commensurate graphene/hBN region of ~10 nm in size, which is surrounded by a strained boundary [19]. All unit cells are characterized by A/B sublattice asymmetry of the same sign, giving a preferred chirality over the entire structure [19,20,25]. For our typical $n$ <10$^{11}$ cm$^{-2}$, Fermi wavelengths are larger than 100nm, which exceeds the superlattice periodicity by a factor of ~10. This large wavelength/period ratio renders contributions from spatially varying couplings insignificant.

We will show below that the observed nonlocality features are consistent with bulk topological currents expected for a gapped Dirac spectrum. The mechanism by which Berry curvature generates topological currents can be elucidated by the semiclassical equations of motion [5]

$$\hbar \dot{k} = eE + ev(k) \times B, \quad v(k) = \frac{1}{\hbar}\frac{\partial \epsilon(k)}{\partial k} + \dot{k} \times \Omega(k) \quad (1)$$

where $\Omega$ is the Berry curvature density and $v(k)$ the group velocity of Bloch electrons. The Lorentz force term $ev \times B$ describes the conventional Hall effect. Berry curvature gives rise to an 'anomalous velocity', $\dot{k} \times \Omega(k)$, which is of the same structure as the Lorentz term but in momentum space and leads to transverse currents (Fig. 1A). Such currents may appear in zero $B$ [without breaking TRS that requires $\Omega(k) = -\Omega(-k)$] as long as inversion symmetry is broken, $\Omega(k) \neq \Omega(-k)$, as is the case of our aligned devices with the globally broken sublattice A/B symmetry [19]. For $B=0$, transverse currents arise solely from $\Omega(k)$, yielding Hall-like conductivity $\sigma_{xy}$ given by the sum of Berry fluxes for all occupied states in the Fermi sea [3]

$$\sigma_{xy} = 2\frac{e^2}{h} \int \frac{d^2k}{2\pi} \Omega(k)f(k) \quad (2)$$

with $f(k)$ being the Fermi function and the factor of 2 accounting for spin degeneracy. Because Berry curvature is odd in energy [20], $\sigma_{xy}$ has the same sign for both electrons and holes. This contrasts with the conventional Hall conductivity that is sign-changing under carrier-type reversal. As a result, $\sigma_{xy}$ given by Eq. 2 is less susceptible to smearing by inhomogeneity. This mechanism yields a non-zero $R_{NL}$ whenever the Fermi level is tuned through Berry curvature hot spots. Their extent in energy is given by half the bandgap $\Delta \approx 180$ K, which translates into $n \approx 2 \times 10^{10}$ cm$^{-2}$ and agrees well with the ultra-narrow width of our $R_{NL}$ peaks.

Because of TRS, the electric field generates topological currents (Eq. 1) with opposite transverse components in graphene's two valleys, $K$ and $K'$ (Fig. 1A), to create the charge-neutral VHE, $J_v = J_K - J_{K'} = \sigma_{xy}^v E$, where $\sigma_{xy}^v = 2\sigma_{xy}$. As illustrated in the inset to Fig. 1B, topological currents can result in a VHE conductivity of $\approx 2e^2/h$. In the absence of intervalley scattering, the charge-neutral currents can persist over extended distances and mediate nonlocal electrical signals [24-28]. The resulting nonlocal resistance $R_{NL}$ can be understood as originating from the VHE and a reverse VHE [20], by analogy with nonlocal transport mediated by charge-neutral spin or energy flow [24-28]. Yet, unlike the latter, the VHE-induced nonlocality appears without TRS breaking, that is, at zero $B$. This behavior, as well as the narrow range of $n$ over which $R_{NL}$ is observed, is a telltale sign of bulk topological currents. The analysis outlined above yields the model expression [20]

$$R_{NL} = (w/2\xi)\,(\sigma_{xy}^v)^2 \rho_{xx}^3 \exp(-L/\xi) \quad (3).$$

The peak in $R_{NL}(n)$ can be described by Eq. 3 with no fitting parameters (Fig. 2B).

The measured spatial decay with $\xi \approx 1.0$ μm is consistent with intervalley scattering occurring at graphene edges and/or at atomic-scale defects [20]. The large values of $R_{NL}$ at $L$ of several μm also imply extremely strong topological currents locally, within the path of the applied current. By extrapolating the observed $L$ dependence to $L <1$ μm, Fig. 2C yields $R_{NL}$ ~10 kOhm which, according to Eq. 3, translates into $\sigma_{xy}^v \approx 2e^2/h$ and order-one Hall angles, in agreement with the VHE expected for weak intervalley scattering. Furthermore, similar to classical magnetotransport, changes in the direction of current flow can lead to additional resistivity. For $\sigma_{xy}\rho_{xx}$ ~1, the classical magnetoresistance reaches a value of ~$\rho_{xx}$ when carriers of opposite sign are involved. A valley analogue of this extra resistance may explain anomalous contributions of ~10 kOhm in $\rho_{xx}$, which are observed at short distances $L \approx w$ by using the bend geometry (Supplementary Fig. S7).

Parenthetically, the intrinsic VHE mechanism discussed above, which provides excellent agreement with our experimental results, may coexist with extrinsic VHE mechanisms such as skew scattering and side jumps [6]. While their role in graphene superlattices remains to be examined, we note that such mechanisms also originate from Berry curvature and arise under the same symmetry conditions as the intrinsic contribution.

Finally, we note that sharp changes in $R_{NL}$ with $V_g$ (Figs. 1-2; 130 nm thick dielectric) amount to a transistor-like response with a slope of ≈100 mV/dec, that is, the detected voltage changes by a factor of 10 by varying $V_g$ by ≈100 mV. Although the peaks in $R_{NL}$ broaden with increasing $T$ and disappear above 100 K (because of the relatively small $\Delta$), one can envision electronic devices based on the valley degrees of freedom [29], which would become practical if larger bandgap values are achieved. To explore this further, we fabricated a superlattice device with a short top gate (15 nm dielectric) placed between the current and voltage contacts used for nonlocal measurements (Supplementary Fig. S8). Valley currents in this case could be switched on and off, similar to the case of a field effect transistor, by a gate voltage of ≈10 mV at 20 K [20]. It is feasible to further reduce the thickness of the top gate hBN dielectric down to 2 nm, which would translate into a gate response down to <2 mV/dec at this $T$. Although further analysis is necessary, these results may indicate that sub-threshold slopes better than those achievable for conventional charge-based processing devices [30] are possible.

*Acknowledgments* – This work was supported by the European Research Council, the Royal Society, the National Science Foundation (STC Center for Integrated Quantum Materials, grant DMR-1231319), Engineering & Physical Research Council (UK), the Office of Naval Research and the Air Force Office of Scientific Research.

1. W. Kohn, J. M. Luttinger. Quantum theory of electrical transport phenomena. *Phys. Rev.* **108**, 590- 611 (1957).
2. G. Sundaram, Q. Niu. Wave-packet dynamics in slowly perturbed crystals: Gradient corrections and Berry-phase effects. *Phys. Rev. B* **59**, 14915-14925 (1999).
3. F. D. M. Haldane. Berry curvature on the Fermi surface: Anomalous Hall Effect as a topological Fermi-liquid property. *Phys. Rev. Lett.* **93**, 206602 (2004).
4. D. Xiao, W. Yao, Q. Niu. Valley-contrasting physics in graphene: magnetic moment and topological transport. *Phys. Rev. Lett.* **99**, 236809 (2007).
5. D. Xiao, C. Meng, Q. Niu. Berry phase effects on electronic properties. *Rev. Mod. Phys.* **82**, 1959-2007 (2010).
6. N. Nagaosa, J. Sinova, S. Onoda, A. H. MacDonald, N. P. Ong. Anomalous Hall effect. *Rev. Mod. Phys.* **82**, 1539 (2010).
7. J. Zak. Berry's phase for energy bands in solids. *Phys. Rev. Lett.* **62**, 2747 -2750 (1989).
8. C. L. Kane, E. J. Mele. Quantum spin Hall effect in graphene. *Phys. Rev. Lett.* **95**, 226801 (2005).
9. B. A. Bernevig, T. L. Hughes, S. C. Zhang. Quantum spin Hall effect and topological phase transition in HgTe quantum well. *Science* **314**, 1757-1761 (2006).
10. L. Fu, C. L. Kane, E. J. Mele. Topological insulators in three dimensions. *Phys. Rev. Lett.* **98**, 106803 (2007).
11. R. Roy. Topological phases and the quantum spin Hall effect in three dimensions. *Phys. Rev. B* **79**, 195322 (2007).
12. M. Konig *et al.* Quantum spin Hall insulator state in HgTe quantum wells. *Science* **318**, 766-770 (2007).
13. C. Z. Chang *et al.* Experimental observation of the quantum anomalous Hall effect in a magnetic topological insulator. *Science* **340**, 167-170 (2013).
14. D. Hsieh *et al.* Observation of unconventional quantum spin textures in topological Insulators. *Science* **323**, 919-922 (2009).
15. D. Xiao *et al.* Coupled spin and valley physics in monolayers of MoS2 and other group-VI dichalcogenides. *Phys. Rev. Lett.* **108**, 196802 (2012).
16. K. F. Mak, K. L. McGill, J. Park, P. L. McEuen. The valley Hall effect in $MoS_2$ transistors. *Science* **344**, 1489-1492 (2014).
17. T. Ando, T. Nakaishi, R. Saito. Berry's phase and absence of back scattering in carbon nanotubes. *J. Phys. Soc. Jpn.* **67**, 2857-2862 (1998).
18. B. Hunt *et al.* Massive Dirac fermions and Hofstadter butterfly in a van der Waals heterostructure. *Science* **340**, 1427-1430 (2013).
19. C. R. Woods *et al.* Commensurate-incommensurate transition for graphene on hexagonal boron nitride. *Nature Phys.* **10**, 451-456 (2014).

20. See the accompanying Supplementary Information.
21. L. A. Ponomarenko *et al*. Cloning of Dirac fermions in graphene superlattices. *Nature* **497**, 594-597 (2013).
22. C. R. Dean *et al*. Hofstadter's butterfly and the fractal quantum Hall effect in moiré superlattices. *Nature* **497**, 598-602 (2013).
23. Z. G. Chen *et al*. Observation of an intrinsic bandgap and Landau level renormalization in graphene/boron-nitride heterostructures. *Nature Commun.* (2014). DOI:10.1038/ncomms5461.
24. D. A. Abanin *et al.* Giant nonlocality near the Dirac point in graphene. *Science* **332**, 328-330 (2011).
25. J. C. W. Song, P. Samutpraphoot, L. S. Levitov. Topological Bands in G/hBN heterostructures. arXiv: 1404.4019 (2014).
26. J. Renard, M. Studer, J. A. Folk. Origins of nonlocality near the Dirac point in graphene. *Phys. Rev. Lett.* **112**, 116601 (2014).
27. M. Titov *et al*. Giant magnetodrag in graphene at charge neutrality. *Phys. Rev. Lett.* **111**, 166601 (2013).
28. D. A. Abanin, A. V. Shytov, L. S. Levitov, B. I. Halperin. Nonlocal charge transport mediated by spin diffusion in the spin Hall effect regime. *Phys. Rev. B* **79**, 035304 (2009).
29. A. Rycerz, J. Tworzydlo, C. W. J. Beenakker. Valley filter and valley valve in graphene. *Nature Phys.* **3**, 172 - 175 (2007).
30. S. M. Sze, K. K. Ng. *Physics of Semiconductor Devices*, chapter 6. Wiley, New York (2007).

## Supplementary Information

**#1 Fabrication, measurement procedures and statistics**

Our graphene superlattice devices were fabricated as follows. First, we mechanically exfoliated an hBN crystal (up to 100 nm thick) and deposited it on top of an oxidized Si wafer (90 or 300 nm of $SiO_2$). We made graphene by cleavage on another substrate and positioned it on top of the hBN crystal (with micrometer accuracy) by using dry or dry-peel transfer procedures detailed in refs. S1-S3. Because both hBN and graphene cleave preferentially along their main crystallographic directions, we exploited their long straight edges to align the crystals using a rotating positioning stage under an optical microscope. We estimate our alignment accuracy to be better than 1.5°. We further crosschecked the alignment using atomic force microscopy, allowing us to visualize the moiré pattern and its periodicity in friction and pica-force modes [19]. For encapsulated devices, a second hBN crystal (typically, ~20 nm thick) was placed on top of the G/hBN stack. The top hBN was intentionally misaligned with respect to graphene by ~10° to avoid extra superlattice effects being induced by the top crystal. A portion of the graphene crystal was left without encapsulation to allow the deposition of metal (Au/Ti) contacts. These contacts were defined by electron beam lithography and deposited by the standard thin film evaporation. Finally, electron beam lithography was used again to make a resist mask defining graphene Hall bars that had 6 to 10 contacts. Oxygen plasma etching was employed to make the final device such as shown in Fig. 1B (inset) of the main text. Prior to measurements, we annealed our structures at 300°C in an argon-hydrogen atmosphere to remove remnant contamination [S4].

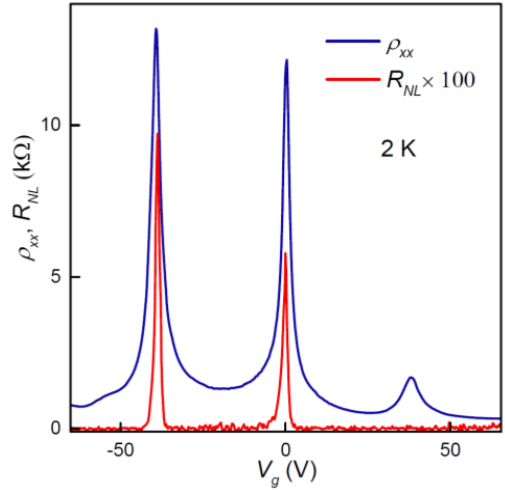

*Figure S1. Resistivity $\rho_{xx}$ and nonlocal resistance $R_{NL}$ in zero magnetic field. The data are for non-encapsulated superlattice device with w ≈1 μm. $R_{NL}$ was measured at separation L ≈5 μm (see the device geometry in Fig. 1 of the main text).*

The measurements were carried out using both DC and low-frequency lock-in techniques (<30Hz) in a variable temperature cryostat. Examples of our raw data are shown in Fig. 1B and Fig. S1. Notably, the observed peaks in $R_{NL}$ are extremely sharp, much sharper than those in $\rho_{xx}$. The local resistivity signal decreases relatively slowly (as approximately $1/V_g$) and, depending on mobility $\mu$, its minimum values between the main and secondary NPs remain large (typically, >300 Ohms for the entire range shown in Fig. S1). This behavior is standard and was previously reported for graphene superlattices [18,21,22]. In contrast, the nonlocal signal rapidly vanishes as $V_g$ is tuned away from the main and hole-side NPs. Within our experimental accuracy of ~0.1 Ohm, $R_{NL}$ is zero everywhere, except in the immediate vicinity of the two NPs.

In total, we measured nonlocal response in 15 aligned devices, 8 of which were encapsulated. All the encapsulated superlattice devices exhibited metallic-like behavior of $\rho_{xx}$ with resistivity saturating at low $T$ to values of <10 kOhms [19,21]. Examples of $\rho_{xx}(T)$ for encapsulated superlattices can be found in refs. 19,21. The activation behavior of conductivity with a low-$T$ insulating state was observed in 4 non-encapsulated devices (Fig. S2), yielding the same bandgap of ~350 K for all of them, in agreement with the values reported in refs. 18,19,23. On the other hand, the nonlocal response was essentially identical for all our superlattice devices, independently of whether they were encapsulated or not.

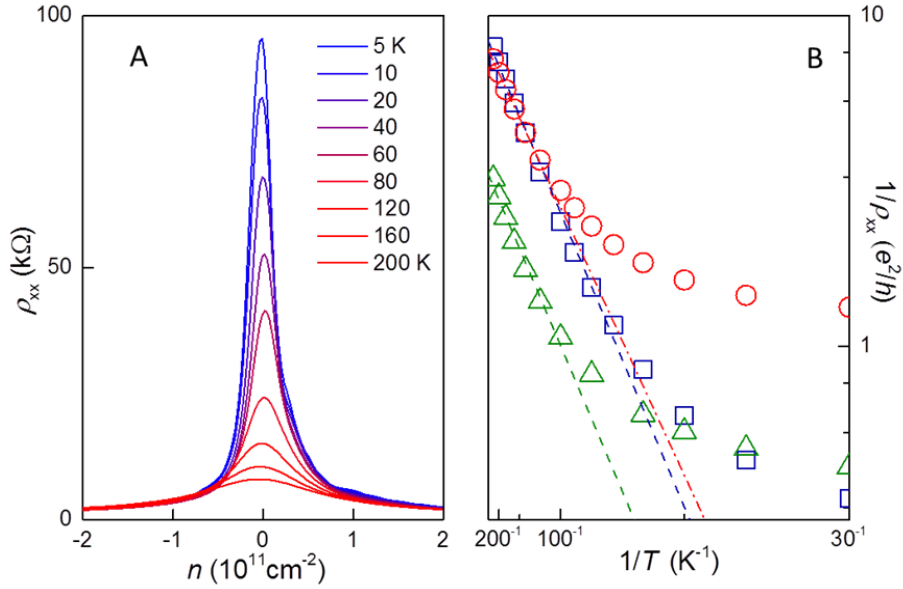

*Figure S2. Non-encapsulated devices often showed the activation behavior. (A) $\rho_{xx}$ reaches ~100 kOhm at 5 K and continues growing with decreasing T. (B) Arrhenius-type plots of $\rho_{xx}$ measured at the main NP for 3 non-encapsulated superlattice devices. Symbols are the experimental data; lines – best fits to the exponential behavior. Within 5% accuracy, the fits yield the same bandgap $2\Delta \approx 350$ K for the three devices. Similar data for another device can be found in ref. 19.*

It is important to note that the absence of an activation behavior in $\rho_{xx}$ does not necessarily imply the absence of a bandgap. In fact, we believe that both encapsulated and non-encapsulated structures are gapped but the gap is obscured in many of them. This can be due to, for example, charge inhomogeneity which would allow current to circumnavigate insulating regions so that, globally, electron transport appears to be non-activated. Alternatively, the metallic-like behavior may appear due to scarcity of midgap impurity states. Such states, if present, pin the Fermi level inside the gap, which is critical for inducing the insulating state and the observation of activated transport. In the absence of midgap states, no activation behavior in $\rho_{xx}$ can be expected because the Fermi level is free to jump between the top of the valence band and the bottom of the conduction band. The difference between the superlattice devices exhibiting activation and non-activation behavior [18,19,21] still remains to be clarified in further studies. Nonetheless, we note that the 4 non-encapsulated devices that exhibited the activation behavior were somewhat lower in electronic quality than those that did not show the transport bandgap. Also, our encapsulated devices were generally higher in quality than non-encapsulated ones. This correlation perhaps points at the scarcity of midgap states as the mechanism that disguises the bandgap. Recent magneto-optical measurements [23], which are less sensitive to charge inhomogeneity and do not require midgap states for detecting the bandgap, provide an alternative way to probe the bandgap and unambiguously prove its presence in graphene superlattices.

Neither charge inhomogeneity nor midgap states are expected to have impact on the VHE and nonlocal response. In the former case (charge inhomogeneity), the nonlocal behavior is preserved upon averaging over density modulation caused by random potential fluctuations. This is seen from Fig. 1 of the main text, which shows that the VHE is given by a non-sign-changing function of density. In the absence of midgap states, we essentially have an ideal semiconductor with no localized states. In the latter case, transport properties are fully controlled by the band states, all of which contribute to the VHE. We therefore expect the VHE to be similar in encapsulated and non-encapsulated graphene superlattices, independent of their temperature dependence in $\rho_{xx}$.

Lastly, we emphasize high reproducibility of our nonlocal measurements. This can be judged from Fig. 2C of the main text, which shows all the available data for different devices for the given temperature. Nominally similar devices are shown in the same color. The relatively small scatter indicates the highly reproducible nature of $R_{NL}$. Moreover, this scatter can be attributed in part to remaining differences in charge inhomogeneity, which can enhance $R_{NL}$ (see Section 6 below). With regard to nonaligned devices, we have investigated several tens of them, and none showed any nonlocality in zero $B$ (several such devices were presented in ref. 24).

**#2 Global gap and optimal conditions for observing the topological currents**

The electronic states responsible for topological currents have a very different character in topological insulators and topological conductors. Let us briefly discuss how these differences impact transport measurements. In topological insulators, the electronic states relevant for electron transport arise due to topologically-protected edge modes. These states lie within the bulk bandgap and determine transport properties if the chemical potential is positioned within the bandgap. States in topological insulator's bulk which may be present due to, e.g., impurities can hamper topological transport by generating currents that short-circuit and obscure the edge state contribution.

In contrast, bulk states play the key role in topological conductors. In this case, currents arising due to Berry curvature can be observed if the chemical potential is positioned within the conduction or valence band, producing a Fermi surface for Bloch states and giving rise to metallic transport in the bulk. The contribution from bulk states in topological conductors is sometimes referred to as the Fermi surface contribution [3]. The key manifestations are distinctly different for the two cases: edge transport and quantized Hall conductivity for topological insulators vs bulk transport and non-quantized Hall conductivity for topological conductors. The Hall conductivity in topological conductors can be nonzero even in zero magnetic field, providing the key experimental signature for transport experiments.

The above general observations translate into important differences in the role of the energy gap for the two kinds of systems. Bandgaps in topological insulators can be thought of as "peepholes" for observing topological transport which is accessed if the Fermi level is positioned within the gap. For topological conductors, on the other hand, a gap opening at an avoided crossing of Bloch bands induces Berry curvature in the bands directly above and below the bandgap, as illustrated in Fig. 1 of the main text. In this case, the optimal condition for observing topological transport is different and involves such doping of the system that the Fermi level is positioned within one of Berry curvature hot spots, above or below the bandgap. Positioning the Fermi level within the gap completely suppresses the contribution of bulk topological currents. Therefore, not only the global gap is inessential for observing valley currents in topological conductors but in fact the transport phenomena due to Berry curvature are best sensed if the Fermi level is positioned slightly away from the gapped region.

Related to this, the role of disorder and charge inhomogeneity is also different in the two cases. For topological insulators, weak disorder helps to pin the Fermi level within the gap, creating optimal conditions for observing edge transport. Strong disorder or inhomogeneity can induce carriers in the bulk, triggering transport mechanisms that obscure edge state transport. In contrast, weak disorder in topological conductors merely adds extra scattering for bulk states but has little effect on the Fermi level position. Additional scattering may change the Hall conductivity value by adding the so-called side-jump and skew-scattering contributions to the intrinsic Hall conductivity arising directly from Berry curvature. Strong inhomogeneity in topological conductors may also create spatial variations in local carrier density, producing locally gapped regions where the carrier density vanishes. However, as long as these regions remain isolated, their presence has little impact on topological currents. At the same time, charge inhomogeneity can strongly obscure the activation behavior in $\rho_{xx}$.

Summing up, the global gap is not required for observing valley currents in topological conductors. Charge inhomogeneity, which often hampers the observation of edge state transport in topological insulators, is expected to have little effect in topological conductors: bulk currents are expected to dominate at moderate inhomogeneities because isolated insulating regions cannot short-circuit the bulk transport. Valley currents in topological conductors are therefore less susceptible to the presence of charge inhomogeneity than those in topological insulators. These points are in agreement with our experimental observations: while the global transport gap has not observed for every superlattice device, all of them featured strong nonlocal response, which is a signature of bulk topological currents.

#3 Berry curvature and valley Hall effect in graphene superlattices
Several contributions to the anomalous Hall effect (valley or spin) have been identified and discussed in literature. The intrinsic contribution, also known as the Fermi sea contribution, essentially counts the Berry flux for the occupied states. In addition, two extrinsic contributions are known, arising due to side-jump and skew-scattering effects [6]. The extrinsic contributions originate from transport and scattering at the Fermi energy. All the contributions, intrinsic and extrinsic, are known to be of a similar magnitude for the well-studied case of the spin Hall effect. We suggest this may also be the case for graphene's VHE. However, it should be noted that the extrinsic mechanisms are in fact also related to Berry curvature and, from a broad symmetry viewpoint, they arise under exactly the same conditions as the Fermi sea contribution. At present, we do not have strong evidence either for or against extrinsic contributions and, because the intrinsic contribution provides good agreement with the experimental results (including their absolute value) we have chosen to concentrate on this mechanism and postpone discussion of possible extrinsic contributions to elsewhere.

The intrinsic contribution arises from a non-zero Berry curvature, $\Omega(k)$. In graphene, $\Omega(k)$ can emerge when inversion symmetry is broken [4,5]. This is because both time reversal symmetry (TRS) and inversion symmetry place separate constraints on $\Omega(k)$. TRS requires that $\Omega(k) = -\Omega(-k)$, and inversion symmetry requires $\Omega(k) = \Omega(-k)$. When both symmetries are respected, $\Omega(k) = 0$. As a result, breaking inversion symmetry is key to creating Berry curvature in the presence of TRS [5], the case of graphene in zero magnetic field.

Experimentally, we induce sublattice asymmetry (thereby breaking inversion symmetry) by aligning graphene on hBN. In these aligned superlattice structures, 'domains' with commensurate stacking between graphene and hBN are formed [19]. The superlattice cells have size of ~10 nm and are surrounded by strained boundaries [19]. There are a number of possibilities for commensurate stacking. For example,
(i) site A in hBN aligned with site A in graphene and site B in hBN with site B in graphene, or
(ii) site A in hBN aligned with site B in graphene and site B in hBN with C in graphene, or
(iii) site A in hBN aligned with site C in graphene and site B in hBN with site A in graphene,
and so on.
Here C denotes the empty position at the center of graphene's hexagonal lattice. The lowest energy configuration out of these commensurate stackings determines the stacking arrangement in all the superlattice cells. Therefore, each of the stacking domains should adopt the same lowest energy configuration. As a result, the uniform A/B sublattice asymmetry across all the domains results in a global inversion symmetry breaking and a global electronic energy gap at the main Dirac point (DP).

It is instructive to consider if stacking frustration of configurations with the same energy can occur. In particular, we note that the following two configurations have the same energy:
(a) site A in hBN aligned with site A in graphene and site B in hBN with site B in graphene
and (b) site A in hBN aligned with site B in graphene whereas site B in hBN aligned with site A in graphene.
Since they cost the same energy, one can imagine a situation where (a) and (b) stackings occur in adjacent domains. However, geometrical constraints do not allow both configurations to occur in adjacent cells in a continuous G/hBN superlattice. This is because while lateral sliding of a graphene sheet from a superlattice cell in configuration (a) can generate neighboring cells in configuration (ii) or (iii), it can never generate configuration (b). The latter would require lattice rotation.

As we now show, global asymmetry between A and B sublattices of the graphene lattice creates hot spots of Berry curvature near the DP, as shown in Fig. 1B inset. We concentrate on the region close to the DP around a single valley which can be described by a gapped Dirac Hamiltonian, $H = v_F \sigma \cdot \mathbf{k} - \Delta \sigma_z$, where $\mathbf{k}$ is the wavevector, $v_F$ is the Fermi velocity and $\sigma$ are Pauli matrices. The gap $\Delta$ reflects the magnitude of global A/B sublattice asymmetry. In G/hBN superlattices, it has been found experimentally [18,19,23] that an energy gap $2\Delta \sim 350$ K opens at the main DP. For this gapped Dirac spectrum, Berry curvature density, $\Omega$, has opposite signs above and below the gap and is given by [4,5]

$$\Omega(k) = \frac{v_F^2 \Delta}{2 [\epsilon(k)]^3}, \quad \epsilon(k) = \pm\sqrt{v_F^2 k^2 + \Delta^2} \qquad \text{(S-1)}$$

where $\epsilon$ is the gapped Dirac fermion energy, and the sign of $\Omega(k)$ depends on the sign of $\Delta$. The equation shows that $\Omega(k)$ decays rapidly with increasing energy, yielding hot spots of Berry curvature near the DP as illustrated in Fig. 1B (inset) of the main text. Interestingly, the odd function $\Omega(k)$ means that $\sigma_{xy}$ (Eq. 2 of the main text) has the same sign for both electrons and holes making it less susceptible to smearing by charge inhomogeneity. Because of TRS, both $\Omega(k)$ and $\sigma_{xy}$ in valley $K$ have opposite signs to those in valley $K'$ leading to an aggregate valley Hall conductivity of $\sigma_{xy}^v = 2\sigma_{xy}$.

Integrating Eq. 2 of the main text, using the Berry curvature of Eq. S-1 and taking $T=0$, we obtain $\sigma_{xy}^v = 2e^2/h$ if the chemical potential lies within the gap. This agrees with our estimate for the observed VHE conductivity, which is obtained by extrapolating the results of our nonlocal resistance measurements to small $L$ (see the main text and below).

Similar hot spots of Berry curvature also appear at the edge of the superlattice Brillouin zone [25], as illustrated in Fig. 1B (inset) of the main text. These occur because of pseudo-spin texture mixing via Bragg scattering off the superlattice potential in G/hBN. In the same way as detailed above, these also produce VHE conductivity and nonlocality.

#4 Modeling nonlocal resistance and its density dependence
Charge-neutral topological currents are detected using nonlocal measurements (Fig. 1 of main text). One can understand the nonlocal signal as a result of the combination of the VHE and a reverse VHE. Indeed, the current applied, $J$, between for example contacts 4 and 8 (Fig. 1 of the main text), induces a valley current $J_v$. This valley current flows perpendicular to $J$ and can create an imbalance in valley populations, $\delta\mu = \mu_K - \mu_{K'}$, between distant contacts 2 and 10 ($\mu_{K,K'}$ are the chemical potentials in the valleys $K$ and $K'$). In turn, $\delta\mu$ leads to a voltage response via the reverse VHE. The VHE and the reverse VHE can be described as

$$J_v = \sigma_{xy}^v E, \qquad E = \frac{\sigma_{xy}^v \rho_{xx}}{2e} \nabla \delta\mu \qquad \text{(S-2)}$$

Accounting for the diffusive dynamics of the valley populations as outlined in the analysis of ref. 28, we obtain $R_{NL}$ described by Eq. 3 in the main text.

Since Berry curvature displays a strong dependence on charge carrier energy, the induced topological currents and, therefore, $R_{NL}$ are also sensitive to carrier density $n$ in G/hBN superlattice devices. To model the $n$ dependence of $R_{NL}$, we first evaluate the density dependence of $\sigma_{xy}^v$ that arises from Berry curvature in a gapped Dirac spectrum. Focusing on the main DP and using Eq. (S-1) and Eq. 2 of the main text, we find

$$\sigma_{xy}^v = \frac{4e^2}{h} \sum_{\pm} \int \frac{d^2 k}{2\pi} \Omega_{\pm}(k) f(\epsilon_{k,\pm} - \mu), \qquad \Omega_{\pm}(k) = \frac{v^2 \Delta}{2\epsilon_{k,\pm}^3}, \qquad \text{(S-3)}$$

where $\epsilon_{k,\pm} = \pm\sqrt{v^2 k^2 + \Delta^2}$ and $\pm$ denote the conduction and valence band contributions, respectively. Taking $T=0$, transforming coordinates from $k \to \epsilon$ and using $v^2 k\, dk = \epsilon d\epsilon$, yields

$$\sigma_{xy}^v = \frac{2e^2 \Delta}{h|\mu|}, \quad |\mu| \geq \Delta \qquad \text{(S-4).}$$

For $\mu$ inside the gap, $\sigma_{xy}^v = \frac{2e^2}{h}$. Finally, to obtain the density dependence of $\sigma_{xy}^v$, we note that for a gapped Dirac spectrum, the carrier density in the conduction band is given by

$$n = 4\int \frac{d^2 k}{(2\pi)^2} f(\epsilon_k - \mu) = \frac{\mu^2 - \Delta^2}{\pi \hbar^2 v^2} \qquad \text{(S-5).}$$

Here the factor of 4 accounts for the spin/valley flavors in graphene. Using Eq. 3 of the main text and approximating $\rho_{xx}(n) \propto 1/(n^2 + \delta n^2)^{1/2}$, we obtain the density dependence of the nonlocal resistance as

$$R_{NL} \propto \sigma_{xy}^{v\,2} \rho_{xx}^3 \propto \frac{n_0}{(n^2+\delta n^2)^{\frac{3}{2}}(|n|+n_0)} \quad , \quad n_0 = \frac{\Delta^2}{\pi \hbar^2 v^2} \quad (S\text{-}6).$$

As illustrated in Fig. 2B of the main text, Eq. (S-6) (black curve) reproduces well the measured nonlocal resistance (red circles) without fitting parameters, by using $\delta n \approx 1.5 \times 10^{10}$ cm$^{-2}$ and $\Delta = 180$ K found from local measurements [18,19,23 and Fig. S2].

#### #5 Temperature dependence of nonlocal signal

The observed nonlocality strongly depends on $T$ as shown in Figs. S3-S4. Above 50 K, $R_{NL}$ decreases exponentially with increasing $T$ and, above 150 K, becomes less than 10 Ohm at the main NP and <1 Ohm at the hole-side NP, even for the shortest $L$ =3.5 µm used in our experiments. The $T$ dependence can be described by the Arrhenius formula exp(-$T^*/T$) over one to two orders of magnitude. This yields the same $T^* \approx 360$ K for both main and hole-side NPs. Both encapsulated and non-encapsulated devices exhibited the same $T^*$ (cf. Figs. S3 & S4). At $T$ below 50 K, $R_{NL}$ tends to saturate to a finite value, presumably because of charge inhomogeneity that, as usual, smears sharp features.

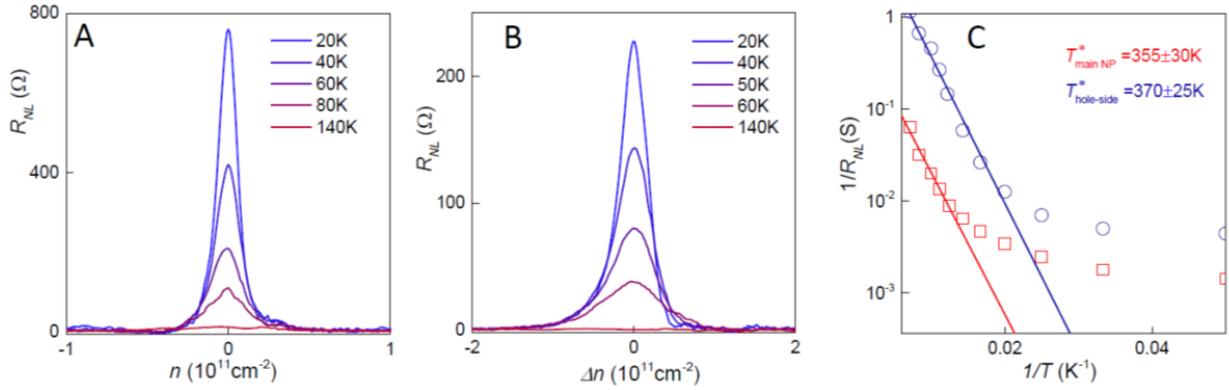

Figure S3. Temperature dependence of nonlocal resistance $R_{NL}$ at the main (**A**) and hole-side (**B**) NPs. Encapsulated G/hBN device; w ≈1 µm and L ≈3.5 µm. The carrier density $\Delta n$ in (B) is counted from the position of the secondary NP. (**C**) Detailed T dependences for the peaks in (A) and (B).

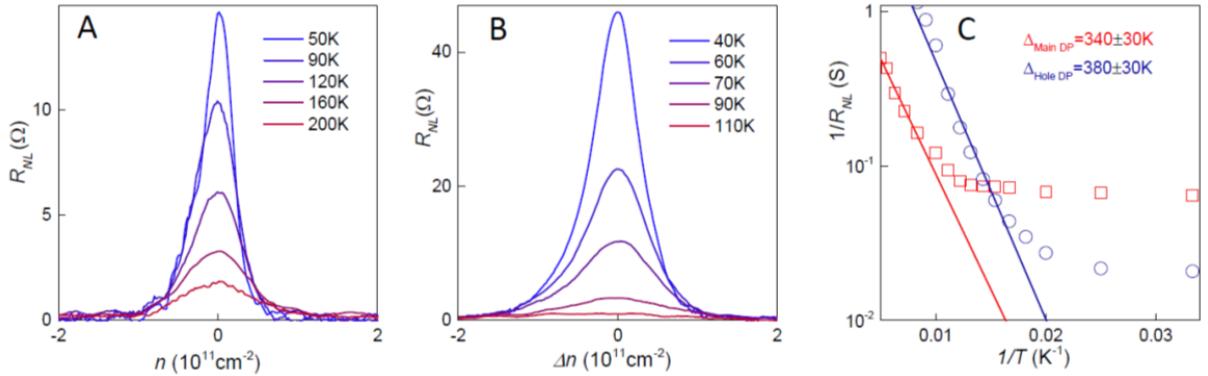

Figure S4. The same as in Fig. S3 but for a non-encapsulated device with w ≈1 µm and L ≈5 µm.

#6 Influence of charge inhomogeneity

Nonlocal resistance is found to depend strongly on charge inhomogeneity $\delta n$. This is shown in Fig. S5. To control $\delta n$, we could sweep gate voltage to progressively larger negative $V_g$, beyond the hole-side NP. This technique exploits the fact that G/hBN has impurities capable of trapping positive charges. The origin of these defects is unknown but we note that they are specific to graphene superlattices. The traps broaden the peaks in $\rho_{xx}$ by up to a factor of 2 but do not significantly reduce $\mu$. Such traps can be neutralized by sweeping to large positive $V_g$ (beyond the electron-side NP) or annealing devices at $T$ above 200°C.

The observed changes in $R_{NL}$ with increasing $\delta n$ indicate that electron-hole puddles can strongly suppress topological currents at the hole-side NP (Fig. S5). This can be attributed to a limited spectral width ($\approx$50 meV) of secondary Dirac spectra [S5,S6]. Indeed, if the energy barrier created by a p-n junction reaches in height this spectral width, the standard Klein tunneling mechanism breaks down, which should result in additional backscattering and suppression of both local and nonlocal transport for secondary Dirac fermions. This scattering at p-n junctions is likely to be the reason for the absence of any detectable $R_{NL}$ at the electron-side NP. Indeed, the secondary spectrum in graphene's conduction band occupies only a narrow energy interval [S5,S6] and, accordingly, the influence of $\delta n$ should be larger for electron-side Dirac fermions. As for the effect of $\delta n$ near the main NP, energy barriers due to p-n junctions remain relatively low with respect to the whole Dirac spectrum, and the observed enhancement of $R_{NL}$ with increasing $\delta n$ (Fig. S5) may indicate that topological currents tend to propagate along p-n junctions. Alternatively, one may invoke side jumps and skew-scattering at charge inhomogeneities, which can also influence $R_{NL}$. However, large values of $R_{NL}$ in the limit of short $L$ (see the main text) seem to indicate that intrinsic rather than extrinsic mechanisms dominate the observed topological currents.

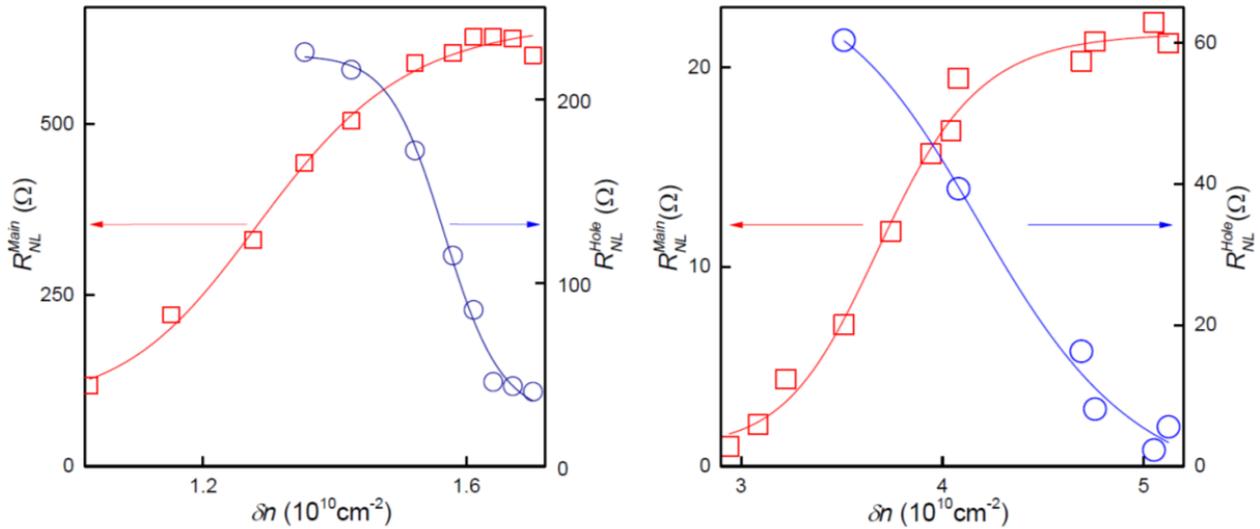

*Figure S5. Nonlocal transport in graphene superlattices as a function of charge inhomogeneity. $\delta n$ is determined from broadening of the resistance peak at the main NP and quantified by fitting it with the dependence $\rho_{xx}(n) \propto 1/(n^2+\delta n^2)^{1/2}$. Left and right panels are for G/hBN devices with L $\approx$ 3.5 µm (encapsulated) and 5 µm (non-encapsulated), respectively.*

#7 Dependence of nonlocal signal on magnetic field

In the case of broken TRS, nonlocal signals can appear due to charge-neutral currents involving spin and energy flow [24-28]. The latter nonlocality has been reported in graphene subjected to strong magnetic fields [24,26]. Therefore, it is essential to show that $R_{NL}$ observed in graphene superlattices is a zero-field phenomenon. To this end, Fig. S6 plots $R_{NL}$ as a function of $B$ applied perpendicular to graphene. One can see that $R_{NL}$ increases with increasing $B$, in agreement with the previously reported mechanisms involving spin/energy flow [24,26,27]. Importantly, the nonlocal signal does not disappear in small $B$ - as had always been the case of the standard graphene devices - but reaches a large finite value in zero $B$. Furthermore, we have observed that the peak in $R_{NL}$ rapidly broadens with increasing $B$ above 0.1 T. This is expected because in the case of spin/energy neutral currents, peaks in $\rho_{xx}(n)$ and $R_{NL}(n)$ exhibit approximately the same width and shape [24-28]. The broadening is further indication that the origin of the ultra-narrow peaks in $R_{NL}$ is different from that of the previously observed nonlocal signals that required broken TRS.

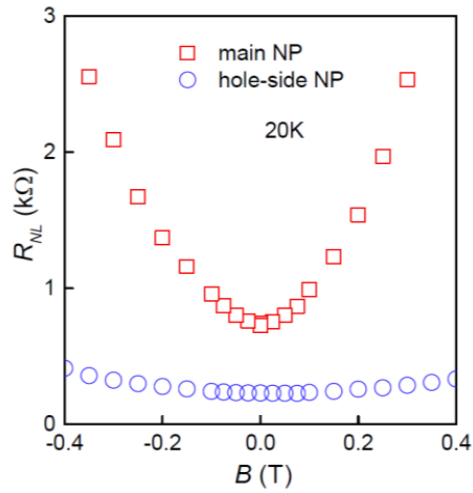

*Figure S6. Nonlocal resistance in G/hBN superlattices does not disappear in zero B. $w \approx 1$ μm; $L \approx 3.5$ μm.*

#8 Anomalous resistivity of graphene superlattices at short distances

We have measured $\rho_{xx}$ in both standard and bend geometries. The standard geometry refers to the case where the electric current $J$ is applied between contacts, for example, 1 and 6 and the voltage drop $V$ is measured between, for example, contacts 3 and 4 (see Fig. 1B of the main text). Then, $\rho_{xx}$ is given by $V/J$ divided by the geometrical factor, $L/w$. The bend geometry describes the case where information about electronic properties is collected from a small area of $\approx w \times w$ (essentially, the Hall cross itself) by applying $J$ through, for example, 4-6 and detecting $V$ at 3-8. In the latter case, $\rho_{xx}$ can be found using the van der Pauw formula, $\rho_{xx} = (V/J)\pi/\ln(2)$. As long as electron transport is diffusive, both geometries yield the same $\rho_{xx}$, independently of the contacts used and independently of whether measurements are done in the bend or standard geometry.

In principle, electron transport in micron-sized graphene devices can enter the ballistic regime such that the mean free path $l$ becomes comparable to $w$. In the latter regime, the van der Pauw formula derived on the basis of the Poisson equation is no longer valid [S1]. However, for $\mu \sim 50,000$ cm$^2$V$^{-1}$s$^{-1}$ and $n < 10^{11}$ cm$^{-2}$, which are relevant to the reported nonlocality, $l < 0.2$μm $\ll w$. Accordingly, both measurement geometries are expected to yield the same result. To demonstrate that this is the case, Figs. S7A-B show $\rho_{xx}(n)$ for a G/hBN device without alignment. There is good agreement between measurements in the bend and standard geometries with only minor variations (10%), caused by imperfections in Hall bars' geometry and/or charge inhomogeneity on the scale of the device's width, $w$.

We find that this conventional behavior breaks down for graphene superlattices as shown in Figs. S7C-D. At high $T$ where the nonlocality is suppressed (Figs S3-S4), differences for $\rho_{xx}(n)$ measured in the two geometries are relatively small. In contrast, we find huge differences between $\rho_{xx}$ in the standard and bend geometries at low $T$ (Fig. S7C). These differences are difficult to accommodate, even assuming that the gap opening can be more pronounced in the bend geometry. This behavior has been seen for all our superlattice devices, both encapsulated and non-encapsulated. Accordingly, we attribute the additional $\rho_{xx}$ contribution detected in the bend geometry to the principal difference between the conditions of Fig. S7C and Figs. S7A,B,D, that is, to the presence of strong topological currents in the former case.

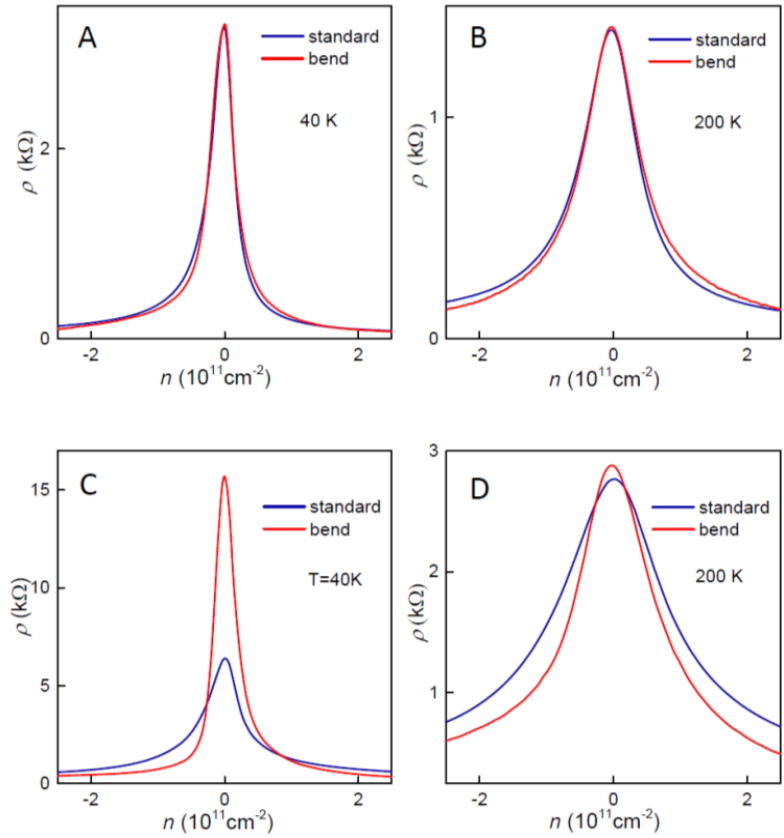

*Figure S7. Anomalous behavior of $\rho_{xx}$ in graphene superlattices at short distances. (**A, B**) – Standard nonaligned G/hBN device. (**C, D**) – G/hBN superlattice under the same conditions. The standard and bend measurements are expected to yield the same $\rho_{xx}(n)$. This is the case of nonaligned graphene devices for all T (A-B). At high T, graphene superlattices also exhibit basic agreement between the two geometries (D). Large differences (factor of 3) are found for superlattice devices at low T where valley currents are strong [for the conditions in (C), $R_{NL}$ reaches 800 Ohm at L ≈3.5 µm].*

The classical model of conductivity for two types of carriers with the same mobility but moving in opposite directions yields magnetoresistance $\Delta\rho \propto \sigma_{xy}^{-2}/\rho_{xx}$ (this corresponds to the case of magnetoresistance in compensated metals). In the case of the valley Hall effect, the same derivation yields additional resistivity $\Delta\rho = \rho_{xx}^{-1}(\sigma_{xy}^v)^{-2}$ in zero $B$. By using $\sigma_{xy}^v \approx 2e^2/h$ as found by extrapolating $R_{NL}$ to $L<w$ (see Fig. 2C of the main text), one can estimate that the expected $\Delta\rho$ is ~ 10kOhm, similar in magnitude to the difference observed between the bend and standard geometries in Fig. S7C. Moreover, the additional contribution detected in the bend geometry (difference between the two curves in Fig. S7C) can be fitted by the same dependence (Eq. S-6) that describes $R_{NL}(n)$ measured at $L \approx 3.5$ µm but multiplied by a factor of $\approx 100$ to reflect the fact that in the bend geometry $L<w$. The difference in the peak values in Fig. S7C is shown by the black star in Fig. 2B of the main text.

#9 Valley-based transistors

Spintronics is a well-established research area that attracts continuous intense interest. Its goal is to exploit the spin degree of freedom for information processing. The basic requirement for this is to be able to create and manipulate a flow of spins [S7]. Despite continuous efforts, it has so far proven difficult to create devices with strong spin polarization and even more challenging to find a way to switch spin currents by gate voltage. As discussed above and in the main text, graphene superlattices allow a high degree of valley polarization with $\sigma_{xy}^v$ values reaching the theoretical maximum of $2e^2/h$. The valley currents can also be switched off by moving the Fermi level away from regions of the electronic spectrum with high Berry curvature. These provide essential elements for so-called 'valleytronics' [29]. To emphasize the possibility to control the valley degree of freedom, we have made a device shown in Fig. S8. In this case, the top gate is much closer (15 nm) to the graphene channel than the back gate. Accordingly, the figure shows that by applying gate voltages of only $\approx 10$ mV, it is possible to suppress valley currents by an order of magnitude. Using thinner gate dielectrics, it should be possible to improve this so-called sub-threshold swing by another order of magnitude.

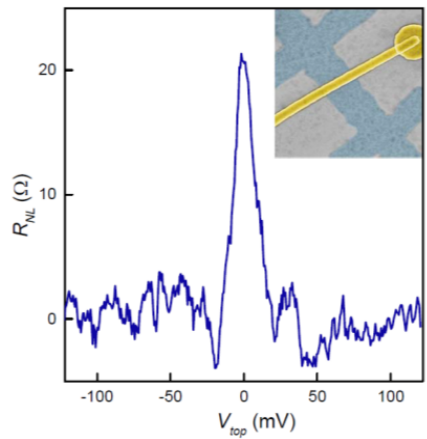

*Figure S8. Valley-based transistor. Nonlocal resistance as a function of top gate voltage $V_{top}$. The inset shows an electron micrograph in false color of the measured superlattice device. G/hBN is in light blue, and the top gate is in gold. $L \approx 4$ µm; $w \approx 1$ µm; T = 20 K. The thickness of the top hBN dielectric is $\approx 15$ nm. The back gate voltage is close to zero but is slightly tuned to reach the main NP.*

S11

**Supplementary references**

S1. A. S. Mayorov *et al*. Micrometer-scale ballistic transport in encapsulated graphene at room temperature. *Nano Lett.* **11**, 2396-2399 (2011).

S2. L. Wang *et al*. One-dimensional electrical contact to a two-dimensional material. *Science* **342**, 614-617 (2013).

S3. A. V. Kretinin *et al*. Electronic quality of graphene encapsulated with different two-dimensional atomic crystals. *Nano Lett.* **14**, 3270-3276 (2014).

S4. S. J. Haigh *et al.* Cross-sectional imaging of individual layers and buried interfaces of graphene-based heterostructures and superlattices. *Nature Mater.* **11**, 764-767 (2012).

S5. J. M. Xue *et al*. Scanning tunnelling microscopy and spectroscopy of ultra-flat graphene on hexagonal boron nitride. *Nature Mater.* **10**, 282–285 (2011).

S6. G. L. Yu *et al*. Hierarchy of Hofstadter states and replica quantum Hall ferromagnetism in graphene superlattices probed by capacitance spectroscopy. *Nature Phys.* **10**, 525-529 (2014).

S7. I. Žutić, J. Fabian, S. Das Sarma. Spintronics: Fundamentals and applications. *Rev. Mod. Phys.* **76**, 323-410 (2004).